\documentclass[aps]{revtex4}
\usepackage[colorlinks=true, pdfstartview=FitV, linkcolor=blue, citecolor=red, urlcolor=magenta, breaklinks=true]{hyperref}
\usepackage{graphicx}  
\usepackage{amsfonts}
\begin{document}
\title{Noncommutative analogue Aharonov-Bohm effect and superresonance}
\author{M. A. Anacleto, F. A. Brito and E. Passos}
\email{anacleto, fabrito, passos@df.ufcg.edu.br}
\affiliation{Departamento de F\'{\i}sica, Universidade Federal de Campina Grande, Caixa Postal 10071, 58109-970 Campina Grande, Para\'{\i}ba, Brazil}
\begin{abstract} 

We consider the idea of modeling a rotating acoustic black hole by an idealized draining bathtub vortex which is a planar circulating flow phenomenon with a sink at the origin. We find the acoustic metric for this phenomenon from a noncommutative Abelian Higgs model. As such the acoustic metric not only describes a rotating acoustic black hole but also inherits the noncommutative characteristic of the spacetime.  We address the issues of superresonance and analogue Aharonov-Bohm (AB) effect in this background. We mainly show that the scattering of planar waves by a draining bathtub vortex leads to a modified AB effect and due to spacetime noncommutativity, the phase shift persists even in the limit where the parameters associated with the circulation and draining vanish. Finally, we also find that the analogue AB effect and superresonance are competing phenomena at a noncommutative spacetime.

\end{abstract}
\maketitle
\pretolerance10000

\section{Introduction}
{ Noncommutative theories have been discussed in the literature by many authors. The inherent nonlocality of these theories leads to the surprising mixture between ultraviolet (UV) and infrared (IR) divergences~\cite{SMRS} which could break the perturbative expansion, lead to loss of unitarity~\cite{GMehen} and violation of Lorentz invariance~\cite{Susskind}.
Noncommutative field theories have also been object of several investigations in planar physics.}

Among several topics in planar physics, the Aharonov-Bohm (AB) effect~\cite{Bohm} is one of the most extensively studied problems. This effect is essentially the scattering of charged particles
by a flux tube and has been experimentally confirmed~\cite{RGC}. {  In  quantum field theory the effect has been simulated, for instance, by using a
nonrelativistic field theory describing bosonic particles
interacting through a Chern-Simons field~\cite{BL}.} It was also found to have analogues in several physical systems such as  in gravitation \cite{FV}, fluid dynamics
\cite{CL}, optics \cite{NNK} and Bose-Einstein condensates \cite{LO} appearing in a vast literature.

It was shown some years ago \cite{Berry} that scattering by a standard vortex leads
to an analogue of the AB effect, determined by a single dimensionless
circulation parameter. More recently, it was shown in~\cite{Dolan} that the scattering of planar waves by a
'draining bathtub' vortex describes a modified AB effect
which depends on two dimensionless parameters associated with the circulation and draining rates \cite{Fetter}.  The
effect was shown to be inherently asymmetric even in the low-frequency limit and leads to novel interference patterns. { In~\cite{ABP2012-1} we 
extended this analysis to a Lorentz-violating background \cite{Bazeia:2005tb} which allows to have persistence of phase shifts even if circulation and draining vanish.}

The purpose of this paper { is to reconsider all the analysis done in~\cite{ABP2012-1} for Lorentz-violating background in order } to investigate the effect of the noncommutativity on the scattering
by a 'draining bathtub' vortex that provides a simple
analogue for the AB effect that naturally occurs in quantum mechanics. Thus, in this work we investigate 
how the AB effect due to a vortex flow is modified by the spacetime noncommutativity. As our results show, we find that there appears small noncommutative correction to the scattering amplitude, which modifies the qualitative and quantitative aspects of the AB effect. 

The noncommutative AB effect has been already studied in the context of quantum mechanics
\cite{FGLR,Chai} and in the quantum field theory approach~\cite{An}.
In~\cite{FGLR} the noncommutative AB effect has been
shown to be in contrast with the commutative situation. 
It was shown that the cross section for the scattering of scalar particles by a
thin solenoid does not vanish even if the magnetic
field assumes certain discrete values. 

In the present calculations,  we apply the acoustic black hole metrics obtained from a relativistic fluid in a noncommutative spacetime~\cite{ABP12} { via the Seiberg-Witten map} and interestingly  we obtain a result  similar to the noncommutative AB effect found in \cite{FGLR}.
A relativistic version of acoustic black holes from the
Lorentz violating Abelian Higgs model has been also presented in~\cite{ABP,ABP11} (see also~\cite{Xian}).
It was found in~\cite{ABP11} that for suitable values of the Lorentz-violating parameter a wider
or narrower spectrum of particle wave function can be scattered with
increased amplitude by the acoustic black hole. This shows how the tuning of such parameter changes the
superresonance phenomenon previously studied in~\cite{SBP}. Thus, the presence of Lorentz-violating background modifies
the quantity of removed energy of the acoustic black hole (see~\cite{MV,Volovik,Unruh} for some reviews and \cite{others} for related issues). 
By computing the superresonance, we conclude that the spacetime noncommutativity also affects the rate of loss of mass of the acustic black hole in a way analogous to what happens in { Lorentz-violating background previous considered in other setups ~\cite{ABP,ABP11,ABP12} }. Thus for suitable values of the spacetime noncommutativity parameter a wider or narrower spectrum of particle wave function can be scattered with increased amplitude by the acoustic black hole.

In our study we shall focus on the differential cross section due to the scattering  of planar
waves by a draining bathtub vortex that leads to
a modified AB effect in a noncommutative spacetime.  
We anticipate that we have obtained a cross section similar to that obtained in~\cite{FGLR} for noncommutative AB effect in quantum mechanics. The result implies that due to the spacetime noncommutativity pattern fringes
can still persist even in the limit where the parameters associated with the circulation and draining go to zero. 
{ In this limit, the noncommutative  background forms a conical defect, which is also responsible for the appearance of the analogue AB effect.}
We also find that the AB effect and superresonance are competing phenomena at a noncommutative spacetime.

The paper is organized as follows. In Sec.~\ref{II} we apply the black hole metrics obtained in the noncommutative  Abelian Higgs model~\cite{ABP12}. We then apply these metrics to a Klein-Gordon-like equation describing sound waves to study the scattering  of planar
waves by a draining bathtub vortex that leads to
a modified AB effect embedded into {\it two types} of a noncommutative spacetime medium by choosing pure magnetic or electric sector.   In Sec.~\ref{conclu} we make our final conclusions.

\section{The Acoustic Metric in Noncommutative Abelian Higgs Model}
\label{II}
{In this section we consider the noncommutative version of the Abelian Higgs model. 
The noncommutativity is introduced by modifying its scalar and gauge sector by replacing the usual product of fields by the Moyal product \cite{SW,SGhosh,rivelles,revnc}.}
Thus, the Lagrangian of the noncommutative Abelian Higgs model in flat space is
\begin{eqnarray}
\label{eqAHM}
\hat{\cal L}&=&-\frac{1}{4}\hat{F}_{\mu\nu}\ast\hat{F}^{\mu\nu} 
+(D_{\mu}\hat{\phi})^{\dagger}\ast D^{\mu}\hat{\phi}+ m^2\hat{\phi}^{\dagger}\ast\hat{\phi}-b\hat{\phi}^{\dagger}\ast\hat{\phi}\ast\hat{\phi}^{\dagger}\ast\hat{\phi},
\end{eqnarray}
{ where the hat indicates that the variable is noncommutative and the $ \ast $-product is the so-called Moyal-Weyl product or star product which is defined in terms of a real antisymmetric matrix $ \theta^{\mu\nu}$ that
parameterizes the noncommutativity of Minkowski spacetime
\begin{eqnarray}
[x^{\mu},x^{\nu}]=i\theta^{\mu\nu}, \quad \mu,\nu=0,1,\cdots,D-1.
\end{eqnarray}
The $ \ast $-product for two fields $f(x)$ and $g(x)$ is given by
\begin{eqnarray}
f(x)\ast g(x)=\exp\left(\frac{i}{2}\theta^{\mu\nu}\partial^{x}_{\mu}\partial^{y}_{\nu}\right)f(x)g(y)\vert_{x=y}.
\end{eqnarray}
In (\ref{eqAHM}) the noncommutative  fields can be expanded in a formal series in $ \theta $. Using the  Seiberg-Witten (SW)  map this expansion can be constructed in terms of the original fields of a commutative theory transforming under the ordinary transformation laws.

Now using the Seiberg-Witten map \cite{SW}, up to the lowest order in the spacetime noncommutative parameter $\theta^{\mu\nu}$, we find} 
\begin{eqnarray}
&&\hat{A}_{\mu}=A_{\mu}+\theta^{\nu\rho}A_{\rho}(\partial_{\nu}A_{\mu}-\frac{1}{2}\partial_{\mu}A_{\nu}),
\nonumber\\
&&\hat{F}_{\mu\nu}=F_{\mu\nu}+\theta^{\rho\beta}(F_{\mu\rho}F_{\nu\beta}+A_{\rho}\partial_{\beta}F_{\mu\nu}),
\nonumber\\
&&\hat{\phi}=\phi-\frac{1}{2}\theta^{\mu\nu}A_{\mu}\partial_{\nu}\phi.
\end{eqnarray}
{ This very useful map allows us to study noncommutative effects in the framework of
commutative quantum field theory. 

Thus the corresponding theory in a commutative spacetime is~\cite{SGhosh}}
\begin{eqnarray}
\label{acao}
\hat{\cal L}&=&-\frac{1}{4}F_{\mu\nu}F^{\mu\nu}\left(1+\frac{1}{2}\theta^{\alpha\beta}F_{\alpha\beta}\right) 
+\left(1-\frac{1}{4}\theta^{\alpha\beta}F_{\alpha\beta}\right)\left(|D_{\mu}\phi|^2+ m^2|\phi|^2-b|\phi|^4\right)
\nonumber\\
&+&\frac{1}{2}\theta^{\alpha\beta}F_{\alpha\mu}\left[(D_{\beta}\phi)^{\dagger}D^{\mu}\phi+(D^{\mu}\phi)^{\dagger}D_{\beta}\phi \right],
\end{eqnarray}
where $F_{\mu\nu}=\partial_{\mu}A_{\nu}-\partial_{\nu}A_{\mu}$ and  $D_{\mu}\phi=\partial_{\mu}\phi - ieA_{\mu}\phi$. As one knows the parameter $\theta^{\alpha\beta}$ is a constant, real-valued antisymmetric $D\times D$- matrix in $D$-dimensional spacetime with dimensions of length squared. For a review see \cite{revnc}.

{Let us briefly review the steps to find the noncommutative acoustic black hole metric from quantum field theory. Firstly, we decompose the scalar field as  $\phi = \sqrt{\rho(x, t)} \exp {(iS(x, t))}$ into the original Lagrangian to find
\begin{eqnarray}
{\cal L}&=&-\frac{1}{4}F_{\mu\nu}F^{\mu\nu}\left(1-2\vec{\theta}\cdot\vec{B}\right)+\tilde{\theta}\left[\partial_{\mu}S\partial^{\mu}S
-2eA_{\mu}\partial^{\mu}S + e^2A_{\mu}A^{\mu} + m^2\right]\rho-\tilde{\theta}b\rho^2
\nonumber\\
&+&\Theta^{\mu\nu}\left[\partial_{\mu}S\partial_{\nu}S-eA_{\mu}\partial_{\nu}S-eA_{\nu}\partial_{\mu}S+e^2A_{\mu}A_{\nu}\right]\rho
+\frac{\rho}{\sqrt{\rho}}\left[\tilde{\theta}\partial_{\mu}\partial^{\mu}+\Theta^{\mu\nu}\partial_{\mu}\partial_{\nu}\right]\sqrt{\rho},
\end{eqnarray}
where $\tilde{\theta}=(1+\vec{\theta}\cdot\vec{B})$, $\vec{B}=\nabla\times\vec{A}$ and $\Theta^{\mu\nu}=\theta^{\alpha\mu}{F_{\alpha}}^{\nu}$. 
In our calculations we consider the case where there is no noncommutativity between space and time, that is $\theta^{0i}=0$ and use $\theta^{ij}=\varepsilon^{ijk}\theta^{k}$, $F^{i0}=E^{i}$ and $F^{ij}=\varepsilon^{ijk}B^{k}$.

Secondly, linearizing the equations of motion around the background $(\rho_0,S_0)$, with $\rho=\rho_0+\rho_1$ and  $S=S_0+\psi$ we find the equation of motion for a linear acoustic disturbance $\psi$ given by a Klein-Gordon equation in a curved space
\begin{eqnarray}
\frac{1}{\sqrt{-g}}\partial_{\mu}(\sqrt{-g}g^{\mu\nu}\partial_{\nu})\psi=0,
\end{eqnarray}
where $g_{\mu\nu}$ just represents the acoustic metrics given explicitly in the examples below. { We should comment that in our previous computation we assumed linear perturbations just in the scalar sector,  whereas the vector field $A_\mu$ remain unchanged.}
 
}

In the following we shall focus on the planar rotating acoustic noncommutative black hole metrics \cite{ABP12} to address the issues of superresonance phenomenon and analogue Aharonov-Bohm effect. For the sake of simplicity, we shall consider {\it two types} of a noncommutative spacetime medium by choosing first pure magnetic sector and then we shall focus on the  pure electric sector.

\pretolerance10000

\subsection{The case $B\neq 0$ and $E=0$}
The acoustic line element in polar coordinates on the noncommutative plane, up to an irrelevant position-independent factor, in the nonrelativistic limit ($v^2\ll c^2$) { was obtained in \cite{ABP12} and is given by }
\begin{eqnarray}
ds^2&=&-[(1-3\theta_{z}B_{z})c^{2}-(1+3\theta_{z}B_{z})(v^2_{r}+v^2_{\phi})]dt^2
-2(1+2\theta_{z}B_{z})(v_{r}dr+v_{\phi}rd{\phi})dt
\nonumber\\
&+&(1+\theta_{z}B_{z})(dr^2+r^2d\phi^2).
\end{eqnarray}
{ where $B_z$ is the magnitude of the magnetic field in the $z$ direction, $ \theta_z $ is the noncommutative parameter, $c$ is the sound velocity in the fluid and $v$ is the fluid velocity.}
We consider the flow with the velocity potential $\psi(r,\phi) = A\ln{r} + B\phi$  whose velocity profile in polar coordinates on the plane is  given by
\begin{eqnarray}
\vec{v}=\frac{A}{r}\hat{r}+\frac{B}{r}\hat{\phi},
\end{eqnarray}
where $B$ and $A$ are the constants of circulation and draining rates of the fluid flow.

Let us now consider the transformations of the time and the azimuthal angle coordinates as follows 
\begin{eqnarray}
&&d\tau=dt+\frac{(1+2\theta_{z}B_{z})Ardr}
{[(1-3\theta_{z}B_{z})c^2r^2-(1+3\theta_{z}B_{z})A^2]},
\nonumber\\
&&d\varphi=d\phi+\frac{ABdr}{r[c^2r^2-A^2]}.
\end{eqnarray}
In these new coordinates the metric becomes
\begin{eqnarray}
\label{ELB}
ds^2\!=\!\tilde{\theta}
\left[-(1-4\Theta)\left(1-\frac{(1+6\Theta)(A^2+B^2)}{c^2r^2}\right)d\tau^2
+\left(1-\frac{(1+6\Theta)A^2}{c^2r^2}\right)^{-1}dr^2
-\frac{2{\tilde\theta}B}{c}d\varphi d\tau+r^2d\varphi^2\right],
\end{eqnarray}
where $\Theta=\theta_{z}B_{z}$ and $\tilde{\theta}=1+\Theta$. The radius of the ergosphere is given by $g_{00}(r_{e}) = 0$, whereas the horizon is given by the coordinate singularity $g_{rr}(r_{h}) = 0$, that is
\begin{eqnarray}
r_{e}=\sqrt{r_{h}^2+\frac{(1+6\Theta)B^2}{c^2}}, \quad r_{h}=\frac{(1+6\Theta)^{1/2}|A|}{c}.
\end{eqnarray}
We can observe from the equation (\ref{ELB}) that for $A > 0$ we are dealing
with a past event horizon, i.e., acoustic white hole and
for $A < 0$ we are dealing with a future acoustic horizon, i.e.,
acoustic black hole.
The metric can be now written in the form
\begin{eqnarray}
g_{\mu\nu}={\tilde\theta}\left[\begin{array}{clcl}
-(1-4\Theta)\left[1-\frac{r_{e}^2}{r^2}\right] &\quad\quad 0& -\frac{{\tilde\theta}B}{cr}\\
0 & \left(1-\frac{r_{h}^2}{r^2} \right)^{-1}& 0\\
-\frac{{\tilde\theta}B}{cr} &\quad\quad 0 & 1
\end{array}\right],
\end{eqnarray}
with inverse $g^{\mu\nu}$
\begin{eqnarray}
\label{metrinv}
g^{\mu\nu}={\tilde\theta}\left[\begin{array}{clcl}
-\frac{(1+4\Theta)}{f(r)} &\quad\quad 0& -\frac{(1+5\Theta)B}{crf(r)}\\
0 & \left(1-\frac{r_{h}^2}{r^2} \right)& 0\\
-\frac{(1+5\Theta)B}{crf(r)} &\quad\quad 0 & \left[1-\frac{r_{e}^2}{r^2}\right]f(r)^{-1}
\end{array}\right],
\end{eqnarray}
where $f(r)=1-\frac{r_{h}^2}{r^2}$. 

We shall now consider the Klein-Gordon equation for a linear acoustic disturbance $\psi(t,r,\phi)$ in the background metric (\ref{metrinv}), i.e.,
\begin{eqnarray}
\frac{1}{\sqrt{-g}}\partial_{\mu}(\sqrt{-g}g^{\mu\nu}\partial_{\nu})\psi=0.
\end{eqnarray}
We can make a separation of variables into the equation above as follows
\begin{eqnarray}
\psi(t,r,\phi)=R(r)e^{i(\omega t-m\phi)}.
\end{eqnarray}
The radial function $R(r)$ satisfies the linear second-order differential equation
\begin{eqnarray}
\label{EQKG}
&&\left[(1+4\Theta)\omega^2-\frac{2(1+5\Theta)Bm\omega}{cr^2}-\frac{m^2}{r^2}\left(1-\frac{r_{e}^2}{r^2}\right)\right]R(r)
+\frac{f(r)}{r}\frac{d}{dr}\left[rf(r)\frac{d}{dr}\right]R(r)=0.
\end{eqnarray}
We now introduce the tortoise coordinate $r^{\ast}$ by using the following equation
\begin{eqnarray}
\frac{d}{dr^{\ast}}=f(r)\frac{d}{dr}, \quad f(r)=1-\frac{r_{h}^2}{r^2}=1-\frac{(1+6\Theta)A^2}{c^2r^2},
\end{eqnarray}
which gives the solution
\begin{eqnarray}
r^{\ast}=r+\frac{\sqrt{(1+6\Theta)}|A|}{2c}\log{\left(\frac{r-\frac{\sqrt{(1+6\Theta)}|A|}{c}}{r+\frac{\sqrt{(1+6\Theta)}|A|}{c}}\right)}.
\end{eqnarray}
Observe that in this new coordinate the horizon $r_{h}=\frac{(1+6\Theta)^{1/2}|A|}{c}$
maps to $r^{\ast}\rightarrow-\infty$ 
while $r\rightarrow\infty$ corresponds to $r^{\ast}\rightarrow+\infty$.
Now, we consider a new radial function, $G(r^{\ast})=r^{1/2}R(r)$ and the modified radial equation obtained from (\ref{EQKG}) is
\begin{eqnarray}
\label{EG}
\frac{d^2G(r^{\ast})}{dr^{\ast2}}+\left[\left((1+2\Theta)\omega-\frac{(1+3\Theta)Bm}{cr^2} \right)^2-V(r)\right]G(r^{\ast})=0, 
\end{eqnarray}
where $V(r)$ is the potential given by
\begin{eqnarray}
V(r)=\frac{f(r)}{4r^2}\left(4m^2-1+\frac{5(1+6\Theta)A^2}{cr^2}\right),
\end{eqnarray}
a form that resembles that given in Refs. \cite{Dolan, ABP2012-1}. 

\subsubsection{Superresonance phenomenon}

The superresonance (analog to the superradiance in black hole physics) is an effect where a spectrum of particle wave function can be scattered with increased amplitude by an acoustic black hole. As a consequence it causes the rate of loss of mass of the acustic black hole. 
In the following we shall compute this effect in the presence of spacetime noncommutativity.

In the asymptotic region ($r^{\ast}\rightarrow\infty$), the equation (\ref{EG}) can be approximately written as follows
\begin{eqnarray}
\label{EG2}
\frac{d^2G(r^{\ast})}{dr^{\ast2}}+\tilde{\omega}^2G(r^{\ast})=0, \quad \tilde{\omega}^2=(1+2\Theta)^2\omega^2,
\end{eqnarray}
which is satisfied by the simple solution
\begin{eqnarray}
\label{sl}
G(r^{\ast})={\cal C}e^{i\tilde{\omega}r^{\ast}}+{\cal R}e^{-i\tilde{\omega}r^{\ast}}\equiv G_{A}(r^{\ast}).
\end{eqnarray}
Notice that the first term in equation (\ref{sl}) corresponds to ingoing wave and the second term to the reflected wave, so that ${\cal R}$ is the reflection coefficient as in usual studies of potential scattering.
The Wronskian of the solutions (\ref{sl}) can be computed to give
\begin{eqnarray}
{\cal W}(+\infty)=-2i\tilde{\omega}(1-|{\cal R}|^2).
\end{eqnarray}
Now, near the horizon region ($r^{\ast}\rightarrow-\infty$), we have
\begin{eqnarray}
\frac{d^2G(r^{\ast})}{dr^{\ast2}}+\left(\tilde{\omega}-m\tilde{\Omega}_{H}\right)^2G(r^{\ast})=0,
\end{eqnarray}
where, $\tilde{\Omega}_{H}=\Omega_{H}(1-3\Theta)$ and $\Omega_{H}=Bc/A^2$ is the angular velocity of the acoustic black hole. We suppose that just the solution identified by ingoing wave is physical, so that
\begin{eqnarray}
G(r^{\ast})={\cal T}e^{i\left(\tilde{\omega}-m\tilde{\Omega}_{H}\right)r^{\ast}}\equiv G_{H}(r^{\ast}).
\end{eqnarray}
The undetermined coefficient ${\cal T}$ is the transmission coefficient of our one dimensional Schroedinger problem. Now the Wronskian of the solution is
\begin{eqnarray}
{\cal W}(-\infty)=-2i\left(\tilde{\omega}-m\tilde{\Omega}_{H}\right)|{\cal T}|^2
\end{eqnarray}
Because both equations are approximate solutions of the asymptotic limit of the modified radial equation, the Wronskian is constant and then
${\cal W}(+\infty)={\cal W}(-\infty)$. 
Thus, we obtain the reflection coefficient
\begin{eqnarray}
\label{refle}
|{\cal R}|^2=1-\left(\frac{\tilde{\omega}-m\tilde{\Omega}_{H}}{\tilde{\omega}}\right)|{\cal T}|^2.
\end{eqnarray}
For frequencies in the interval $0 < \tilde{\omega} <m\tilde{\Omega}_{H}$ the reflectance is always larger than unit, which implies in the superresonance phenomenon~\cite{SBP,ABP11}.
Here $m$ is the azimuthal mode number and $\Omega_{H}=Bc/A^2$ is the angular velocity of the usual Kerr-like acoustic black hole. Furthermore, the interval of frequencies can be wider or narrower depending on the noncommutative parameter $\Theta=B_z\theta_z$. As a consequence, the tuning of the noncommutative parameter changes the rate of loss of mass of the acustic black hole.


\subsubsection{Analogue Aharonov-Bohm effect}
Let us now consider the analogue Aharonov-Bohm effect by considering the scattering of a monochromatic planar wave of frequency $\omega$ given in the form
\begin{eqnarray}
\psi(t,r,\phi)=e^{-i\omega t}\sum_{m=-\infty}^{\infty}R_{m}(r) e^{im\phi}/\sqrt{r},
\end{eqnarray}
such that far from the vortex, the function $\psi$ can be written in terms of the sum of a plane wave and a scattered wave, i.e.,
\begin{eqnarray}
\psi(t,r,\phi)\sim e^{-i\omega t}(e^{i\omega x}+f_{\omega}(\phi)e^{i\omega r}/\sqrt{r}),
\end{eqnarray}
where $e^{i\omega x}=\sum_{m=-\infty}^{\infty}i^mJ_{m}(\omega r) e^{im\phi}$ and $J_{m}(\omega r)$ 
is a Bessel function of the first kind. The scattering amplitude $f_{\omega}(\phi)$ has the
partial-wave representation
\begin{eqnarray}
f_{\omega}(\phi)= \sqrt{\frac{1}{2i\pi\omega}}\sum_{m=-\infty}^{\infty}(e^{2i\delta_{m}}-1) e^{im\phi},
\end{eqnarray}
and the phase shift is defined as
\begin{equation}
e^{2i\delta_{m}}=i(-1)^m\frac{{\cal C}}{{\cal R}}.
\end{equation}
In order to compute the phase shift, at some level of approximation, let us first rewrite the equation (\ref{EG}) in terms of a new function $X(r)=f(r)^{1/2}G(r^{\ast})$, that is
\begin{eqnarray}
&&\frac{d^2X(r)}{dr^{2}}+ \left(-\frac{3r_{h}^2}{f(r)r^4}+\frac{r_{h}^4}{f(r)^2r^6}\right)X(r)
+\left[\left((1+2\Theta)\omega-\frac{(1+3\Theta)Bm}{cr^2} \right)^2-V(r)\right]
\frac{X(r)}{f^2(r)}=0, 
\end{eqnarray}
that written as a power series in $1/r$, we have
\begin{eqnarray}
\frac{d^2X(r)}{dr^{2}}
+\left[\tilde{\omega}^2-\frac{(4\tilde{m}^2-1)}{4r^2}+U(r)\right]X(r)=0, 
\end{eqnarray}
where $\tilde{m}^2=m^2+2am-2b^2$, $a=\tilde{\omega}(1+3\Theta)B$, $b=\tilde{\omega}(1+3\Theta)A$ and
\begin{eqnarray}
U(r)=\frac{(a^2-b^2)m^2-4b^2am+2b^2+3b^4}{\tilde{\omega}^2r^4}+\frac{b^2(2a^2-b^2)m^2-6b^4am+3b^4+4b^6}{\tilde{\omega}^4r^6}
+O(\tilde{\omega}^{-6}r^{-8}),
\end{eqnarray}
$a$ and $b$ being parameters that describe the coupling to {\it circulation} and {\it draining}, respectively. Now applying the approximation formula
\begin{eqnarray}
\delta_{m}\approx \frac{\pi}{2}(m-\tilde{m})+\frac{\pi}{2}\int^{\infty}_{0}r[J_{\tilde{m}}(\widetilde{\omega}r)]^2U(r)dr,
\end{eqnarray}
and using $|m|\gg\sqrt{a^2+b^2}$, we obtain~\cite{Dolan,ABP2012-1}
\begin{eqnarray}\label{modes-m}
\delta_{m}\cong- \frac{a\pi}{2}\frac{m}{|m|}+\frac{3\pi(a^2+b^2)}{8|m|}-\frac{5a\pi(a^2+b^2)}{8m^2}\frac{m}{|m|},
\end{eqnarray}
with the isotropic mode $m = 0$ giving the imaginary phase shift
\begin{eqnarray}\label{modes-0}
\delta_{m=0}=\frac12{i\pi b}.
\end{eqnarray}
Thus, by using Eqs. (\ref{modes-m}) and  (\ref{modes-0}), to lowest order in $a$ and $b$, we can compute the differential scattering cross section (with units of length) that is given by
\begin{eqnarray}
\frac{d\sigma_{ab}}{d\phi}=|f_{\omega}(\phi)|^2\cong \frac{\pi}{2\tilde{\omega}}\frac{[a\cos(\phi/2)-b\sin(\phi/2)]^2}{\sin^{2}(\phi/2)}.
\end{eqnarray}
For $b = 0$ (the nondraining limit), we have the vortex result of Fetter \cite{Fetter}
\begin{eqnarray}
\label{eqvortex}
\frac{d\sigma_{vortex}}{d\phi}= \frac{\pi a^2}{2\tilde{\omega}}\cot^2(\phi/2)=\frac{(1-2\Theta)\pi^2 a^2}{2\pi\omega}\cot^2(\phi/2),
\end{eqnarray}
that should be compared with the Aharonov-Bohm (AB) effect in the small angle or small coupling limits
\begin{eqnarray}
\frac{d\sigma_{AB}}{d\phi}=\frac{1}{2\pi\tilde{\omega}}\frac{\sin^2(\pi a)}{\sin^{2}(\phi/2)}
=\frac{(1-2\Theta)}{2\pi\omega}\frac{\sin^2(\pi a)}{\sin^{2}(\phi/2)}.
\end{eqnarray} 
For small $\theta$ and small angle $\phi$, the Eq.~(\ref{eqvortex}) becomes
\begin{eqnarray}
\frac{d\sigma_{vortex}}{d\phi}=\frac{(1-2\Theta)\pi^2 a^2}{2\pi\omega}\left[\frac{4}{\phi^2}-\frac{2}{3}+\frac{\phi^2}{60}+O(\phi^3)\right].
\end{eqnarray}
In the present case, if the circulation $a\to0$ the differential scattering cross section vanishes. This is not necessary true for other noncommutative sectors. In the next section we shall show this explicitly.

Taking into account the superresonance phenomenon described above let us relate it to the AB effect for small $\theta$. We see that the range of frequencies can be written as
\begin{eqnarray}
0<\omega<m(1-5\theta)\Omega_H,
\end{eqnarray}
and the differential cross section (\ref{eqvortex}) written in terms of the circulation parameter $a$ and frequencies $\tilde{\omega}$ is 
\begin{eqnarray}
\label{eqvortex-k}
\frac{d\sigma_{vortex}}{d\phi}\simeq \frac{\pi}{2}(1+8\theta)\omega\cot^2(\phi/2)\lesssim  \frac{\pi}{2}(1+3\theta)m\Omega_H\cot^2(\phi/2).
\end{eqnarray}
Notice that the range of scattering frequencies is narrowing (widening) while the differential cross section  is increasing (decreasing) with the noncommutativity strength. This means that AB effect is favored by the scattering of low-frequency waves --- this is also implicit in the approximation (\ref{modes-m}) --- whereas the superresonance becomes small. In summary, AB effect and superresonance are competing phenomena at a noncommutative spacetime. 

\subsection{The case $B=0$ and $E\neq 0$}

In the present subsection we repeat the previous analysis for  $B=0$ and $E\neq 0$. As in the earlier case we take the acoustic line element obtained in \cite{ABP12}, in polar coordinates on the noncommutative plane, up to first order in $\theta$, in the `non-relativistic'  limit ($v^2\ll c^2$), given by
\begin{eqnarray}
\label{am}
ds^2&=&\left(1-\frac{3}{2}\theta\vec{\cal E}\cdot\vec{v}\right)
\left\{-[c^{2}-(v^2_{r}+v^2_{\phi}+\theta{\cal E}_{r}{v}_{r}+\theta{\cal E}_{\phi} {v}_{\phi})]dt^2
-2\left(v_{r}+\frac{\theta{\cal E}_{r}}{2}\right)drdt\right.
\nonumber\\
&-&\left.2\left(v_{\phi}+\frac{\theta{\cal E}_{\phi}}{2}\right)rd{\phi}dt
+(1-\theta{\cal E}_{r}{v}_{r}-\theta{\cal E}_{\phi} {v}_{\phi})(dr^2+r^2d\phi^2)\right\}.
\end{eqnarray}
where $ \theta\vec{\cal E}=\theta\vec{n}\times\vec{E} $, 
$ \theta{\cal E}_{r}= \theta(\vec{n}\times\vec{E})_{r}$ , 
$  \theta{\cal E}_{\phi}=\theta (\vec{n}\times\vec{E})_{\phi}$ and { $E$ is the magnitude of the electric field}.  Let us now consider the transformations of the time and the azimuthal angle coordinates as follows
\begin{eqnarray}
&&d\tau=dt+\frac{\tilde{v}_{r}dr}{(c^2-\tilde{v}_{r}^2)},
\nonumber\\
&&d\varphi=d\phi+\frac{\tilde{v}_{\phi}\tilde{v}_{r}dr}{r(c^2-\tilde{v}_{r}^2)}.
\end{eqnarray}
where we have defined $ \tilde{v}_{r}= v_{r}+\frac{\theta{\cal E}_{r}}{2}$ and $\tilde{v}_{\phi}=v_{\phi}+\frac{\theta{\cal E}_{\phi}}{2} $.
Now, we consider the flow with the velocity potential $\psi(r, \varphi) = A\ln{r}+ B\varphi$ whose velocity profile in polar coordinates on the plane is given by
$\vec{v}=\frac{A}{r}\hat{r}+\frac{B}{r}\hat{\phi}$.
Therefore, in these new coordinates the metric becomes
\begin{eqnarray}
\label{me}
ds^2\!&=&\!\left(1-\frac{3\theta{\cal E}_{r}A}{2r}-\frac{3\theta{\cal E}_{\phi} B}{2r}\right)\left\{
-\left[1-\frac{(A^2+B^2+\theta{\cal E}_{r}Ar+\theta{\cal E}_{\phi}Br)}{c^2r^2}\right]d\tau^2\right.
\nonumber\\
&+&\left.\left(1-\frac{\theta{\cal E}_{r}A}{r}-\frac{\theta{\cal E}_{\phi} B}{r}\right)\left[
\left(1-\frac{A^2+\theta{\cal E}_{r}Ar}{c^2r^2}\right)^{-1}dr^2+r^2d\varphi^2\right]
-2\left(\frac{B}{cr}+\frac{\theta{\cal E}_{\phi}}{2c}\right)rd\varphi d\tau\right\},
\end{eqnarray}
The radius of the ergosphere is given by $g_{00}(\tilde{r}_{e}) = 0$, whereas the horizon is given by the coordinate singularity $g_{rr}(\tilde{r}_{h}) = 0$, that is
\begin{eqnarray}\label{horizonR}
\tilde{r}_{e}=\frac{\theta{\cal E}_{r}A+\theta{\cal E}_{\phi} B}{2c^2}\pm\frac{1}{2}
\sqrt{\frac{(\theta{\cal E}_{r}A+\theta{\cal E}_{\phi} B)^2}{c^4}+4r^2_{e}}, 
\quad\quad \tilde{r}_{h_\pm}=
\frac{\theta{\cal E}_{r}A}{2c^2}\pm r_{h}\sqrt{1+\frac{(\theta{\cal E}_{r})^2}{4c^2}},
\end{eqnarray}
where $ r_{e}=\sqrt{(A^2+B^2)/c^2 }$  and $ r_{h}=|A|/c $ are the radii of the ergosphere and the horizon in the usual case. For $ \theta=0 $, we have  $\tilde{r}_{e}={r}_{e}$ and $\tilde{r}_{h}={r}_{h}  $.

Therefore, we can rewrite Eq. (\ref{me}) in the limit of small $\theta$ as follows
\begin{eqnarray}\label{MMA}
ds^2=\left(1-\frac{\theta \xi c^2}{2r}\right)\left[ -\left( 1-\frac{R^2_e}{r^2}\right)d\tau^2
+\left(1-\frac{R^2_{h}}{r^2}\right)dr^2-\frac{2\cal{B}}{cr}rd\varphi d\tau+r^2d\varphi^2\right],
\end{eqnarray}
where
\begin{eqnarray}
&&R^2_{e}=\left(1+\frac{\theta \xi c^2}{r}\right)r^2_{e} +\theta \xi(1-c^2)r, \quad\quad 
R^2_{h}=r^2_{h}+\frac{\theta{\cal E}_{r}Ar}{c^2}, \quad \quad 
{\cal B}=\frac{2B+\theta {\cal E}_{\phi}}{2}+\frac{\theta B\xi c^2}{r},
\nonumber\\
&& \xi=({\cal E}_{r}A+{\cal E}_{\phi}B)/c^2.
\end{eqnarray}
The components of the metric are
\begin{eqnarray}
g_{\mu\nu}=\left(1-\frac{\theta \xi c^2}{2r}\right)\left[\begin{array}{clcl}
\left[1-\frac{R_{e}^2}{r^2}\right] &\quad\quad 0& -\frac{{\cal B}}{cr}\\
0 & \left(1-\frac{R_{h}^2}{r^2} \right)^{-1}& 0\\
-\frac{{\cal B}}{cr} &\quad\quad 0 & 1
\end{array}\right],
\end{eqnarray}
and	its inverse $g^{\mu\nu}$ reads
\begin{eqnarray}
\label{metrinv}
g^{\mu\nu}=\left(1-\frac{\theta \xi c^2}{2r}\right)\left[\begin{array}{clcl}
-f(r)^{-1} &\quad\quad 0& -\frac{{\cal B}}{crf(r)}\\
0 & \left(1-\frac{R_{h}^2}{r^2} \right)& 0\\
-\frac{{\cal B}}{crf(r)} &\quad\quad 0 & \left[1-\frac{R_{e}^2}{r^2}\right]f(r)^{-1}
\end{array}\right],
\end{eqnarray}
where $f(r)=1-\frac{R_{h}^2}{r^2}$. Now, the radial function $R(r)$, as in the previous case, satisfies the linear second order differential equation
\begin{eqnarray}
\label{eqm}
&&\left[\omega^2-\frac{2{\cal B}m\omega}{cr^2}-\frac{m^2}{r^2}\left(1-\frac{R_{e}^2}{r^2}\right)\right]R(r)
+\frac{f(r)}{r}\frac{d}{dr}\left[rf(r)\frac{d}{dr}\right]R(r)=0.
\end{eqnarray}

 We now just as in the previous case we introduce the tortoise coordinate $r^{\ast}$ by using the following equation
\begin{eqnarray}
\frac{d}{dr^{\ast}}=f(r)\frac{d}{dr}, \quad f(r)=1-\frac{R_{h}^2}{r^2}=1-\frac{r^2_{h}}{r^2}-\frac{\theta{\cal E}_{r}A}{c^2r},
\end{eqnarray}
which gives the solution
\begin{eqnarray}
r^{\ast}=r+\frac{\tilde{r}_{h_+}^2}{\tilde{r}_{h_+}-\tilde{r}_{h_-}}\ln{|r-\tilde{r}_{h_+}|}-\frac{\tilde{r}_{h_-}^2}{\tilde{r}_{h_+}-\tilde{r}_{h_-}}\ln{|r-\tilde{r}_{h_-}|}.
\end{eqnarray}
Observe that in this new coordinate the horizons $\tilde{r}_{h_\pm}$ defined in (\ref{horizonR})
map to $r^{\ast}\rightarrow\mp\infty$ 
while $r\rightarrow\infty$ corresponds to $r^{\ast}\rightarrow+\infty$. 

Let us now focus on the coordinate running from the outer horizon to infinity, i.e., from $r=r_{h_+}$ to $r=\infty$.

Again, we introduce  a new radial function $G(r^{\ast})=r^{1/2}R(r)$ and the modified radial equation (\ref{eqm}), becomes
\begin{eqnarray}
\label{EGG}
\frac{d^2G(r^{\ast})}{dr^{\ast2}}
+\left[\omega^2-\frac{2{\cal B}m\omega}{cr^2}+\frac{m^2\eta^2 }{r^4}-V(r)\right]G(r^{\ast})=0, 
\end{eqnarray}
where,
\begin{eqnarray}
V(r)&=&\frac{f(r)}{4r^2}\left(4m^2-1+\frac{5r_{h}^2}{r^2}+\frac{3\theta {\cal E}_{r}A}{c^2r}\right),
\\
\eta^2&=& \left(1+\frac{\theta \xi c^2}{r}\right) \left( \frac{B^2}{c^2}-\frac{\theta {\cal E}_{r}Ar}{c^2} \right)+\theta \xi(1-c^2)r +\frac{\theta\xi c^2r^2_{h}}{r}.
\end{eqnarray}

\subsubsection{Superresonance phenomenon}

Here, in order to compute the superresonance phenomenon, we follow the same analysis done in the pure magnetic case. 
Thus, in the asymptotic region ($r^{\ast}\rightarrow\infty$), the equation (\ref{EGG}) can be approximately written as
\begin{eqnarray}
\frac{d^2G(r^{\ast})}{dr^{\ast2}}+\omega^2G(r^{\ast})=0,
\end{eqnarray}
which solution is 
\begin{eqnarray}
\label{sl}
G(r^{\ast})={\cal C}e^{i{\omega}r^{\ast}}+{\cal R}e^{-i{\omega}r^{\ast}}\equiv G_{A}(r^{\ast}).
\end{eqnarray}
Now, near the horizon region ($r^{\ast}\rightarrow-\infty$), we have
\begin{eqnarray}
\frac{d^2G(r^{\ast})}{dr^{\ast2}}+\left[\left(\omega-m\tilde{\Omega}_{H}\right)^2
- 2m\omega\left(\frac{\theta{\cal E}_{\phi}}{2B}+\frac{\theta\xi c^3}{A} \right) +m^2\left( \frac{\theta{\cal E}_{r}A^2}{cB^2}+\frac{\theta\xi c^3}{A}+\frac{\theta\xi cA}{B^2}\right)\right]G(r^{\ast})=0,
\end{eqnarray}
where, $\tilde{\Omega}_{H}=\Omega_{H}\left(1-\frac{\theta {\cal E}_r}{c}\right)$ and $\Omega_{H}=Bc/A^2$ is the angular velocity of the acoustic black hole.

For the sake of simplicity, we assume there exist relationships among the parameters $A,B,{\cal E}_r,{\cal E}_\phi$, e.g., 
\begin{eqnarray}
A = -\frac12(B-\sqrt{B^2-2})Bc, \qquad {\cal E}_\phi = -\frac{2{\cal E}_rB(4A+c)}{4B^2+1},
\end{eqnarray}
 such that we simply have
\begin{eqnarray}
\frac{d^2G(r^{\ast})}{dr^{\ast2}}+\left(\omega-m\tilde{\Omega}_{H}\right)^2G(r^{\ast})=0.
\end{eqnarray}
Now we repeat the analysis of the previous case. Again, we suppose that just the solution identified by ingoing wave is physical, so that
\begin{eqnarray}
G(r^{\ast})={\cal T}e^{i\left({\omega}-m\tilde{\Omega}_{H}\right)r^{\ast}}\equiv G_{H}(r^{\ast}).
\end{eqnarray}
The undetermined coefficient ${\cal T}$ is the transmission coefficient of our one dimensional Schroedinger problem. Now the Wronskian of the solution is
\begin{eqnarray}
{\cal W}(-\infty)=-2i\left({\omega}-m\tilde{\Omega}_{H}\right)|{\cal T}|^2
\end{eqnarray}
Just as in the previous case, because both equations are approximate solutions of the asymptotic limit of the modified radial equation, the Wronskian is constant and then
${\cal W}(+\infty)={\cal W}(-\infty)$. 
Thus, we obtain the reflection coefficient
\begin{eqnarray}
\label{refle}
|{\cal R}|^2=1-\left(\frac{{\omega}-m\tilde{\Omega}_{H}}{{\omega}}\right)|{\cal T}|^2.
\end{eqnarray}
Again, for frequencies in the interval $0 < {\omega} <m\tilde{\Omega}_{H}$ the reflectance is always larger than unit, which implies in the superresonance phenomenon.  Furthermore, the interval of frequencies can be wider or narrower depending on the noncommutative parameter relationship ${\cal E}_r\theta$. Thus, just as in the pure magnetic case, the tuning of the noncommutative parameter affects the rate of loss of mass of the acustic black hole.

\subsubsection{Analogue Aharonov-Bohm effect}

As in the previous case, in order to compute the phase shift, at some level of approximation, let us first rewrite the equation 
(\ref{EGG}) in terms of a new function $X(r) = f(r)^{1/2}G(r^*)$, that is
\begin{eqnarray}
&&\frac{d^2X(r)}{dr^{2}}+ \left[\frac{-3\left(r_{h}^2+\theta{\cal E}_{r}Ar/3c^2\right)}{f(r)r^4}+\frac{\left(r_{h}^2+\theta{\cal E}_{r}Ar/2c^2\right)^2}{f(r)^2r^6}\right]X(r)
\nonumber\\
&&+\left[\omega^2-\frac{2{\cal B}m\omega}{cr^2}+\frac{m^2\eta^2 }{r^4}-V(r)\right]
\frac{X(r)}{f^2(r)}=0, 
\end{eqnarray}
and then written as a power series in $1/r$, we have
\begin{eqnarray}
\frac{d^2X(r)}{dr^{2}}
+\left[{\omega}^2+\frac{2\theta b\omega{\cal E}_{r}}{r}-\frac{(4\tilde{m}^2-1)}{4r^2}+U(r)\right]X(r)=0, 
\end{eqnarray}
where we define the parameters $\tilde{m}^2=m^2+(2a+\theta{\cal E}_{\phi})m-2b^2$, $a=\omega B$, $b=\omega A$ and the function
\begin{eqnarray}
U(r)&=&\frac{({a}^2-b^2)m^2-4b^2am+2b^2+3b^4}{{\omega}^2r^4}+\frac{b^2(2{a}^2-b^2)m^2-6b^4am+3b^4+4b^6}{{\omega}^4r^6}
\nonumber\\
&+&\theta\left[-\frac{(m^2-1/2)b{\cal E}_{r}+4bam{\cal E}_{r}+m^2{\cal E}_{r}b+2ma({\cal E}_{r}b+{\cal E}_{\phi}a)}{\omega r^3}
-\frac{2b^2m{\cal E}_{\phi}}{\omega^2 r^4}\right.
\nonumber\\ 
&+&\left.\frac{(3b^3-4ab^3m){\cal E}_{r}-4a^2b^2m{\cal E}_{\phi}+m^2(a^2b{\cal E}_{r}+a^3{\cal E}_{\phi}+b^3{\cal E}_{r}+b^2a{\cal E}_{\phi})
-2m^2b^3{\cal E}_{r}}
{\omega^3 r^5}\right.
\nonumber\\
 &-&\left.\frac{3b^4m{\cal E}_{\phi}}{\omega^4 r^6}\right]+O(\tilde{\omega}^{-5}r^{-7}).
\end{eqnarray}
As in the previous case, by applying again the approximation formula
\begin{eqnarray}
\delta_{m}\approx \frac{\pi}{2}(m-\tilde{m})+\frac{\pi}{2}\int^{\infty}_{0}r[J_{\tilde{m}}({\omega}r)]^2U(r)dr,
\end{eqnarray}
we obtain
\begin{eqnarray}
\delta_{m}&\cong&- \frac{(2a\pi+\pi\theta{\cal E}_{\phi}+4b\theta{\cal E}_{r})}{4}
\frac{m}{|m|}+\left[\frac{\pi(3a^2+3b^2+2a{\cal E}_{\phi}\theta)}{8}
-(a^2\theta{\cal E}_{\phi}
+ab\theta{\cal E}_{r})\right]\frac{1}{|m|}
\nonumber\\
&+&\left[-\frac{(2a+\theta{\cal E}_{\phi})\pi(5a^2+5b^2
+2a{\cal E}_{\phi}\theta)}{16}+\frac{\theta(7a^3{\cal E}_{\phi}
-7b^3{\cal E}_{r}+7a^2b{\cal E}_{r}+ab^2{\cal E}_{\phi})}
{3}\right]\frac{1}{m^2}\frac{m}{|m|}.
\end{eqnarray}
Thus, to lowest order in $a$, the differential scattering cross section with $b = 0$ and for small $\theta$, is
\begin{eqnarray}
\label{eqvortex2}
\frac{d\sigma_{vortex}}{d\phi}= \frac{\pi (a+\theta{\cal E}_{\phi}/2)^2}{2{\omega}}\cot^2(\phi/2).
\end{eqnarray}
Starting from this equation one may find the same conclusion of the previously analysis that AB effect and superresonance are competing  phenomena. 

Computing the differential scattering cross section for small angles $\phi\neq 0$, we have
\begin{eqnarray}
\frac{d\sigma_{vortex}}{d\phi}=\frac{\pi^2(a^2+a\theta{\cal E}_{\phi}+\theta^2
{\cal E}^2_{\phi}/4)}{2\pi\omega}\left[\frac{4}{\phi^2}-\frac{2}{3}+\frac{\phi^2}{60}+O(\phi^3)\right].
\end{eqnarray}
Now, if the $a = 0$, the differential cross section at small angles is dominated by
\begin{eqnarray}
\frac{d\sigma_{vortex}}{d\phi}=\theta^2{\cal E}^2_{\phi}\frac{\pi}{2\omega\phi^2}.
\end{eqnarray}
Note that, contrarily to the usual Aharonov-Bohm effect, in the noncommutative case the differential
scattering cross section is different from zero when $a=0$. 
Our results are qualitatively in agreement with that obtained in~\cite{FGLR}, for the AB effect in the context of noncommutative quantum mechanics. This correction vanishes in the limit $\theta\rightarrow 0$  so that no singularities are generated.
This correction ($\sim\theta^2$) due to effect of spacetime noncommutativity may be relevant at high energy physics or even in condensed matter physics since noncommutativity can manifest there.  In this case the theory of the electrons in a constant magnetic field, projected to the lowest Landau level, is naturally thought of as a noncommutative field theory \cite{revnc}.  Our result shows that pattern fringes can appear even when $a=0$, unlike the commutative case. One can make some estimative of the aforementioned effect by estimating  $\theta$ following similar calculations already known in the literature \cite{casana,mckellar}. 

{ One can understand this effect as follows. In the limit of circulation $a=\omega B$ and draining  $b=\omega A$ vanishes then for nonzero $\omega$, we automatically have $A=B=r_h=r_e=0$ such that the metric (\ref{MMA}) 
simply becomes the metric of a conical defect 
\begin{eqnarray}
\label{am-2}
ds^2=(-d\tau^2+dr^2+r^2d\varphi^2-2\frac{\theta{\cal E}_{\phi}}{2c}rd\varphi d\tau),
\end{eqnarray}
that is
 \begin{eqnarray}
\label{am-22}
ds^2=\left(-d\tilde{\tau}^2+dr^2+r^2\left(\frac{{\theta}{\cal E}_{\phi}}{2c}\right)^{-1}d\tilde{\varphi}^2-2rd\tilde{\varphi} d\tilde{\tau}\right),
\end{eqnarray}
where $\tilde{\tau}=\left(\frac{{\theta}{\cal E}_{\phi}}{2c}\right)^{1/2}\tau$,  $\tilde{\varphi}=\left(\frac{{\theta}{\cal E}_{\phi}}{2c}\right)^{1/2}\varphi$ with angle deficit $\delta=2\pi\left(1-\left(\frac{{\theta}{\cal E}_{\phi}}{2c}\right)^{-1/2}\right)$. Thus, even though there is no vortex in the above limit, the noncommutative background forms a conical defect, which is responsible for the appearance of the analogue AB effect in an even more interesting  way. 
 }

\section{Conclusions}
\label{conclu}

In this paper we have considered the superresonance phenomenon and analogue Aharonov-Bohm effect  due to an idealized draining bathtub vortex in a noncommutative spacetime whose metric is the metric of a noncommutative acoustic black hole obtained in noncommutative Abelian Higgs model \cite{ABP12}.  We show there exist both  superresonance phenomenon and analogue Aharonov-Bohm (AB) effect in this background. In the superresonance effect the range of frequencies of scattering waves where the superresonance occurs can be wider or narrower depending on relations involving the parameter that controls the spacetime noncommutativity. On the other hand, the scattering of planar waves by a draining bathtub vortex leads to a modified AB effect and due to spacetime noncommutativity the phase shift persists even in the limit where the parameters associated with the circulation and draining vanish, as previously showed in Aharonov-Bohm effect in the context of noncommutative quantum mechanics ~\cite{FGLR}. {Furthermore, even though there is no vortex in certain limits, the noncommutative background can still forms a conical defect, which is responsible for the appearance of the analogue AB effect in an even more interesting way. 
}
Finally, we also have found that AB effect and superresonance are competing phenomena at a noncommutative spacetime.

\acknowledgments

We would like to thank CNPq, CAPES and PNPD/PROCAD -
CAPES for partial financial support.

\end{document}